\documentclass{aip-cp}

\usepackage[numbers]{natbib}
\usepackage{rotating}
\usepackage{graphicx}

\begin{document}

\title{Theory Motivation For Exotic Signatures: Prospects and Wishlist for Run II}

\author[aff1]{Daniel Stolarski\corref{cor1}}

\affil[aff1]{Theory Division, Physics Department, CERN, CH-1211 Geneva 23, Switzerland}
\corresp[cor1]{Corresponding author: daniel.stolarski@cern.ch}

\begin{flushright}  
{\normalsize{  
CERN-PH-TH-2015-300
  }   
  }
\end{flushright}
\vspace*{-1.5cm}  
{\let\newpage\relax\maketitle}

\begin{abstract}
Here I give some motivations for exotic signatures to search for at Run II of the LHC, focusing on displaced phenomena. I will discuss signatures arising from various different kinds of models including theories of dark matter and those with exotic decays of the Higgs. 
\end{abstract}

\section{WHAT IS EXOTICS?}

I was given the charge of motivating exotics searches in Run II without being given a definition of exotics. ATLAS and CMS have exotics groups, and looking at the searches within these groups, the common theme appears to be signals that do not appear in supersymmetry (SUSY). Yet, nearly every exotic search can be rewritten in terms of a SUSY model. For example, leptoquark searches look for the operator
\begin{equation}
{\cal L} = LQ \; q \;\ell  
\label{eq:leptoquark}
\end{equation} 
where $q$ and $\ell$ are the Standard Model (SM) quark and lepton, and $LQ$ is the particle being searched for. But if we simply relabel $LQ \rightarrow \tilde{d}$, then the operator in Equation~\ref{eq:leptoquark} is exactly the operator that appears in the $R$-parity violating superpotential in SUSY: $W=\lambda' L \, Q\,D$.  

Another example of exotics secretly being SUSY is a search for a diboson resonance which has received a great deal of attention lately~\cite{Aad:2015owa}. Supersymmetric extensions of the Standard Model often have a problem of predicting a Higgs mass much lighter than the observed value. In the MSSM for example, at tree level $m_h \leq m_Z$. One of the most elegant solutions to this problem is ``non-decoupling $D$-terms''~\cite{Batra:2003nj,Maloney:2004rc} which require a new gauge force and therefore new heavy vector bosons. These bosons must couple to the Higgs to raise the Higgs mass, and therefore should also couple to $W$ and $Z$ bosons, so such a model of non-decoupling $D$-terms could fit a potential excess in a diboson search~\cite{Collins:2015wua}. See also~\cite{Petersson:2015rza} for another SUSY explanation of such an excess presented at LHCP.

Therefore, I use a very different and much more experimentally based definition of exotics: new experimental objects that cannot be produced in the SM. In this talk, I will consider a subset of this definition and focus on displaced signatures, those arising from decays of \textit{long-lived} exotic particles somewhere in the detector but away from the interaction point. Finally, I will give a disclaimer that even with this narrow focus, I only give a few examples of scenarios that give rise to these signatures.

The simplest motivation for exotics is that it could be there, and if it is, we do not want to miss any new physics discoveries. Furthermore, the majority of searches for new physics are looking under the lamppost, namely they are looking for theories for which perturbation theory can be used to make precise predictions. Yet, nature need not be so kind as to allow us to use perturbation theory, so we need to explore as many types of theories as possible. Below I give some more concrete motivations for exotics, but we must keep a broad perspective in our experimental searches.

\section{DARK MATTER}

Cosmological observations give extremely strong evidence for the existence of dark matter and for it making up about one quarter of the energy budget of the universe. Yet, the particle physics properties of dark matter are still completely unknown. A well studied example of a dark matter candidate is the weakly interacting massive particle (WIMP), but searches at both the LHC and in direct detection experiments have thus far only placed limits on WIMP scenarios. Therefore, considering other scenarios for dark matter is extremely well motivated.

\subsection{Freeze-In}

In the WIMP scenario, dark matter is in equilibrium with the SM thermal bath until the temperature drops below its mass when the annihilation of dark matter freezes out and sets the present day abundance. An alternative scenario is the so-called freeze-in mechanism~\cite{Hall:2009bx}, where dark matter is never in thermal equilibrium with the SM, but couples very weakly so that the SM thermal bath slowly leaks energy into the dark sector. This small coupling sets the abundance, and it was recently shown~\cite{Co:2015pka} that the size of this small parameter naturally implies long-lived particles that decay within the detector at LHC experiments. 

In these models, the process is $B \rightarrow A_{\rm SM} X$ where $X$ is dark matter with a mass of at most 100 GeV, but it could be orders of magnitude lighter. $B$ is a new state with large couplings to the SM but small couplings to dark matter that sets the freeze-in abundance, so the $B$ field is naturally long lived. $A_{\rm SM}$ can be virtually any SM state such as $h,\;Z,\;\ell^+\ell^-,\; q\bar{q},\;\gamma,$..., so these models give a motivated scenario to search for virtually any SM state originating in any or all of the sub-detector regions.

\subsection{Asymmetric Dark Matter and Emerging Jets}

The ratio of the energy density of dark matter to baryons in our universe is about five, but it could have been orders of magnitude larger or smaller. In the WIMP paradigm, there is no explanation for why these energy densities are similar. An alternative is asymmetric dark matter: the number density of dark matter is controlled by the fact that there are more dark matter particles than anti-dark matter particles, much like the baryon asymmetry of our universe. This is an old idea~\cite{Nussinov:1985xr} reviewed in~\cite{Petraki:2013wwa}. If the same physics controls the dark matter and baryon asymmetry, then you naturally get that the number density of dark matter and baryons is comparable. But in most of the models that do this, the mass of the dark matter is a free parameter that needs to be set to be similar to the proton mass by hand, and therefore these models do not fully explain the coincidence of dark matter and baryons energy density. 

The mass of the proton is explained by dimensional transmutation, so a theory of dark matter that has a QCD-like sector whose confinement scale is similar to that of QCD could then explain this coincidence, and such a model was presented in~\cite{Bai:2013xga}. In such a theory, there is a whole zoo of hadrons in the dark sector that will also have GeV scale masses. 

\begin{figure}[h]
  \centerline{\includegraphics[width=180pt]{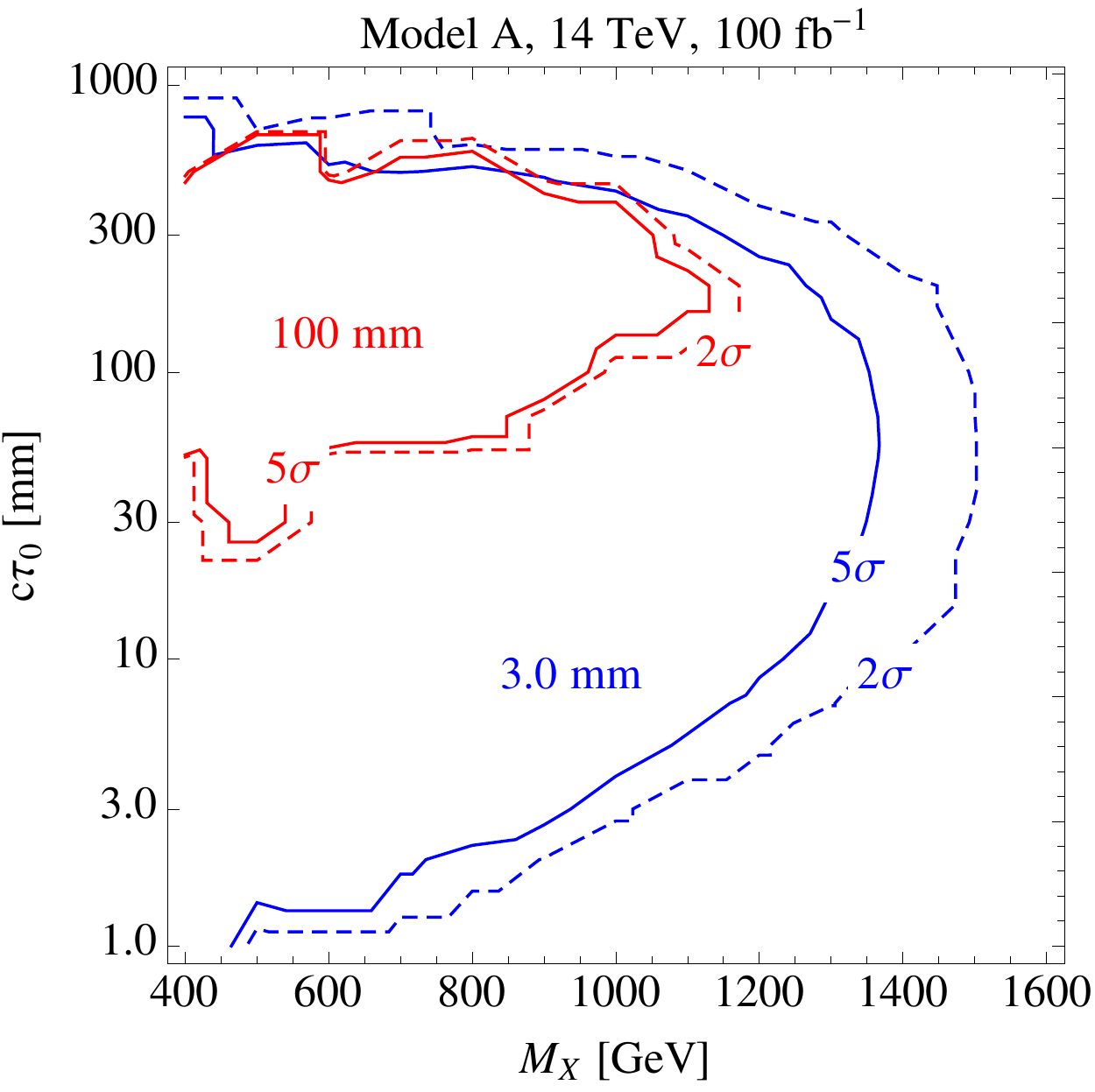}}
  \caption{(Figure 10 from~\cite{Schwaller:2015gea}) Discovery reach for the emerging jets scenario presented in~\cite{Schwaller:2015gea}. The horizontal axis is the mediator mass which controls the production cross section, and the vertical is the dark pion lifetime.}
  \label{fig:reach}
\end{figure}

If there exists a heavy (TeV scale) mediator that couples to SM fields and dark quarks, something that automatically happens in the model of~\cite{Bai:2013xga}, then one could produce dark quark pairs at the LHC. Because the mediator is much heavier than the confinement scale, this process would result in jets of dark sector hadrons. The existence of the mediator also causes the dark pions to decay back into SM fields, and the natural length scale of this decay is $\mathcal{O}$(cm). Therefore, the jet which starts out completely invisible at short distance slowly appears with each dark hadron decaying in a different place and creating a different displaced vertex. We have termed this structure \textit{emerging jets}~\cite{Schwaller:2015gea}, and the signatures at the LHC are quite spectacular with a discovery potential for mediators well into the TeV scale, as shown in Figure~\ref{fig:reach}.

\section{OTHER MOTIVATIONS FOR DISPLACED SIGNATURES}

\subsection{Exotic Higgs Decays}

The usual gauge hierarchy, the quadratic sensitivity of the Higgs mass to high scale physics is most problematic for the top quark loop. The hierarchy problem is solved and the top loop is cancelled by fermionic (scalar) top partners in theories of composite Higgs (supersymmetry). The dominant production mechanism of these partners at the LHC occurs because they are coloured, but they do not need to be! Twin Higgs~\cite{Chacko:2005pe} (folded SUSY~\cite{Burdman:2006tz}) models have uncoloured fermionic (scalar) top partners that can still cancel the SM top loop. 

\begin{figure}[h]
  \centerline{\includegraphics[width=220pt]{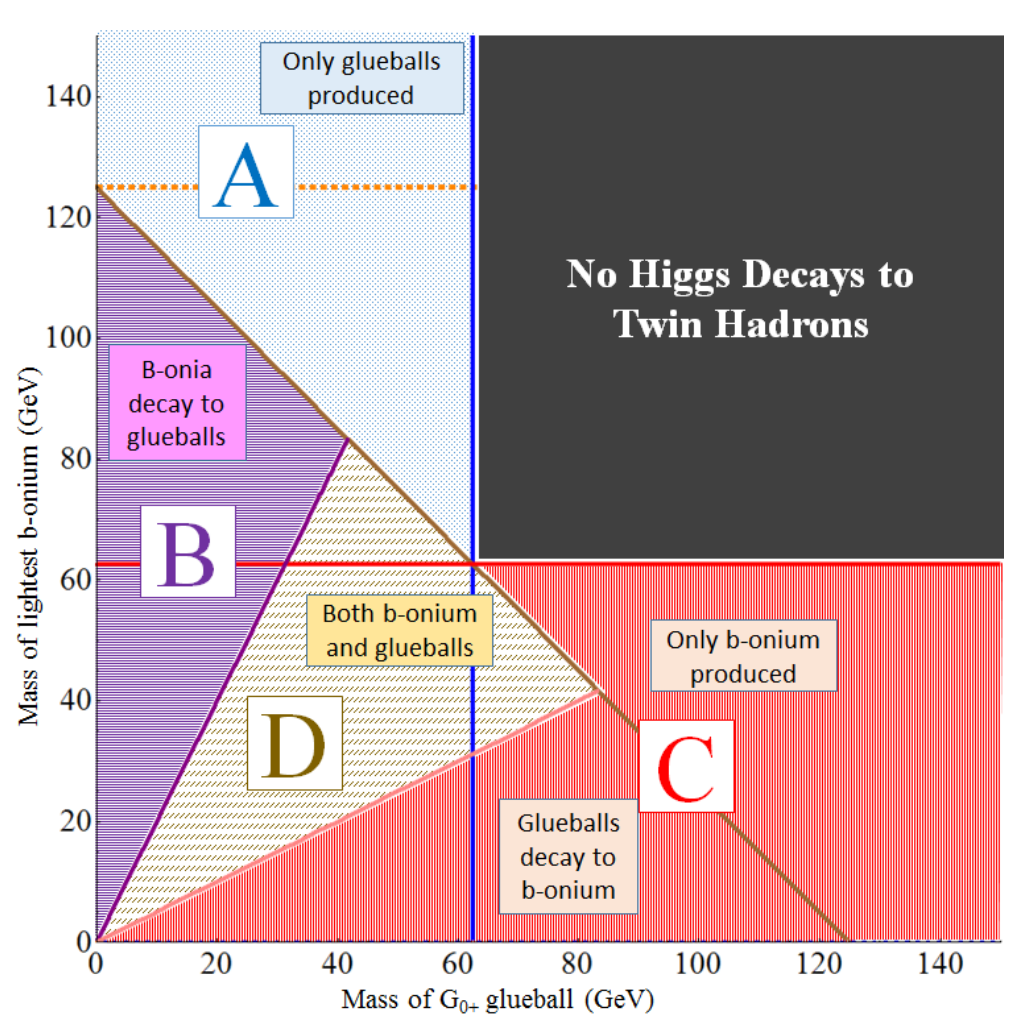}}
  \caption{(Figure 5 from~\cite{Craig:2015pha}) A description of the different kinds of exotic Higgs decays in twin Higgs models as a function of the mass of the lightest twin glueball and the lightest twin bottomonium state. }
  \label{fig:higgsdec}
\end{figure}

In order for these mechanisms to work, there still needs to be a colour factor in the loop, so many of these models have a twin colour gauge group that confines at the GeV scale. Some models have signatures that are similar to emerging jets discussed above, but the mediator can be the SM Higgs, motivating searches for exotics Higgs decays with displaced vertices. The signatures can be quite rich depending on the spectrum of the different confined twin states, and some of the possibilities are detailed in Figure~\ref{fig:higgsdec}~\cite{Craig:2015pha}. Other possibilities are studied in~\cite{Curtin:2015fna}, and other models that give displaced Higgs decays as well as their prospects for the LHC are given in~\cite{Csaki:2015fba}.

\subsection{Quirks}

Most of the models described in this talk involve positing an additional confining gauge group near the GeV scale. The confinement scale of a gauge theory, $\Lambda$ is exponentially sensitive to high-scale parameters, so one can easily imagine such a theory where the confinement scale is many orders of magnitude lower, corresponding to a macroscopic length scale. If we further add fermions charged under this confining group that are also charged under the SM and at the TeV scale, so called quirks~\cite{Kang:2008ea}, then this very innocuous modification in theory space leads to extremely dramatic signatures at the LHC. 

\begin{figure}[h]
  \centerline{\includegraphics[width=180pt]{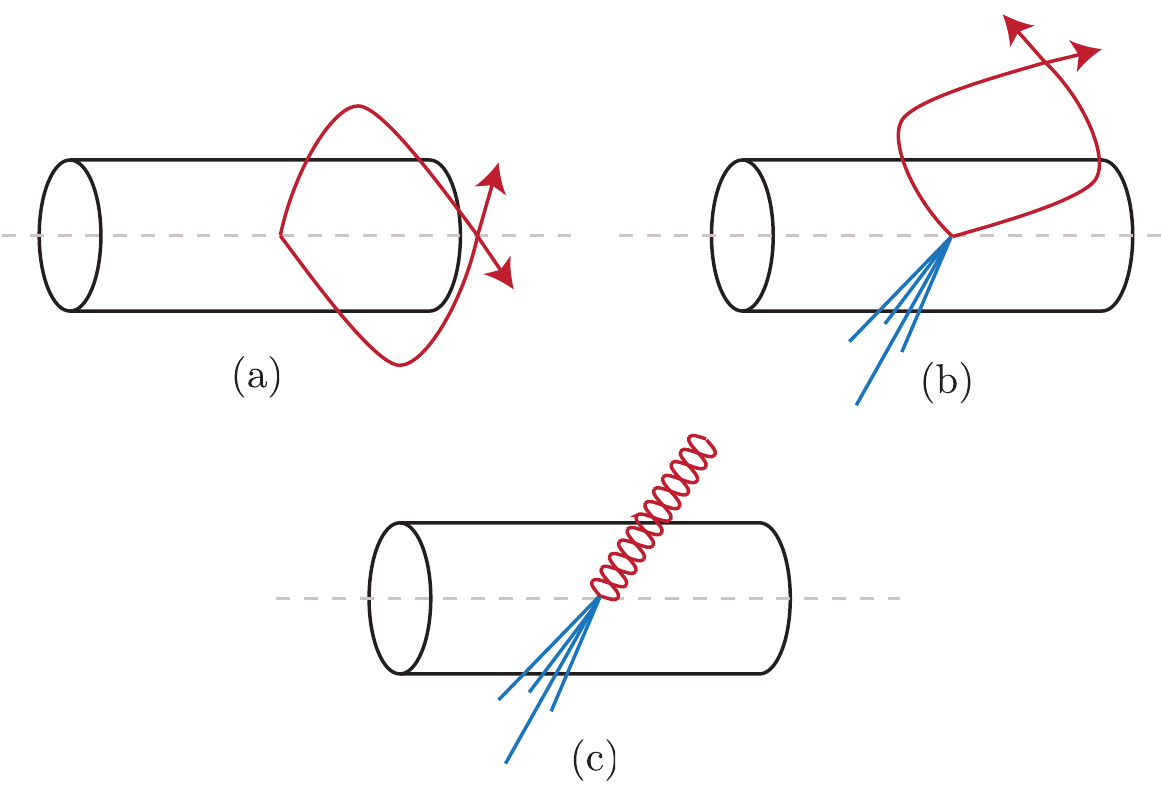}}
  \caption{(Figure 5 from~\cite{Kang:2008ea}) A pictorial representation of some possible quirk signatures.  }
  \label{fig:quirks}
\end{figure}

These quirks will be produced at the LHC and fly apart until they are separated by a distance $\Lambda^{-1}$, and then they will be pulled back together by the confining string. Therefore there will be charged particles taking very strange oscillating paths through an LHC detector. Some of the possibilities are shown in Figure~\ref{fig:quirks}. 

\subsection{A Note on Triggers}

Both ATLAS and CMS have triggers designed to look specifically for different kinds of exotic signatures, and these are an important component of the search program. The LHC, however, is a hadron machine, so jets are quite plentiful. Therefore, standard triggers that look for jets or even leptons can be have a reasonable efficiency for various new physics models. This strategy is already used in mono-jet searches where the new physics is completely invisible, but this strategy can be generalized. 

In~\cite{Csaki:2015fba}, an excellent example of this was given in models where the Higgs decays to long lived neutrals. They compare the trigger efficiencies of three standard triggers: Vector Boson Fusion (VBF), VBF plus $b$'s, and isolated lepton, to two exotic triggers: those for displaced jets and for trackless jets. While those exotic triggers were designed with this sort of new physics in mind, in many regions of parameter space the standard triggers do as well or better than the exotic ones. That is because the new physics sometimes happens to produced in conjunction with jets or gauge bosons, so the triggers can pick up those objects and leave the search for new physics to the analysis level.

\section{WISHLIST FOR RUN II}

My personal wishlist for exotic searches in Run II:
\begin{itemize}
\item More searches for distinct collider objects such as emerging jets or quirks. 
\item Searches for different SM states originating in different places in the detector. 
\item More general use of triggers including multi-jet and VBF. 
\item Keep searches as model independent as possible, trying not to use the details of any particular model for physics beyond the SM. 
\end{itemize}
With this, if new physics exists in an exotic form, it is more likely to be found by the LHC's search program. 


\section{ACKNOWLEDGMENTS}
I would like to thank the organizers of LHCP for the invitation to give this talk, and particularly Andreas Weiler for his feedback on my talk. 


\nocite{*}
\bibliographystyle{JHEP}%
\bibliography{Exotics}%

\end{document}